\pdfoutput=1
\documentclass[12pt,reqno]{article}
\usepackage{jheppub}
\usepackage{amsmath,amssymb,amsfonts}
\graphicspath{ {./Graphs/} }
\usepackage{float}             
 \usepackage[T1]{fontenc}
\usepackage[utf8]{inputenc}
\usepackage{mathtools}  
\usepackage{blkarray}
\usepackage[usenames,dvipsnames]{xcolor}
\usepackage{epsfig}
\usepackage{epstopdf}
\usepackage{dcolumn}
\usepackage{tikz}
\usetikzlibrary{shapes.geometric, arrows}
\usepackage{upgreek}
\usepackage{setspace}
\usepackage{enumitem}
\usepackage{array,multirow,bigdelim}
\usepackage{cleveref}
\usepackage{bm}

\def\be{\begin{equation}}
\def\ee{\end{equation}}
\def\ba{\begin{eqnarray}}
\def\ea{\end{eqnarray}}

\def\a{\alpha}
\def\b{\beta}

\def\b#1{\overline{#1}}

\def\CP1{\mathbb{CP}^1}
\def\SL2C{\mathrm{SL}(2,\mathbb{C})}

\def\Z2{\mathbb{Z}_2}

\def\su2{{SU(2)}}
\def\eps{{\epsilon}}
\def\a{{\alpha}}

\def\[{\left[}
\def\]{\right]}

\def\L{\Lambda}

\def\s{\sigma}
\def\a{\alpha}
\def\b{\beta}

\def\({\left(}
\def\){\right)}
\def\[{\left[}
\def\]{\right]}

\def\<{\langle}
\def\>{\rangle}

\def\i2{\frac{i}{2}}

\newcommand{\ie}{{\it i.e.}}
\newcommand{\eg}{{\it e.g.}}
\newcommand{\bea}{\begin{eqnarray}}
\newcommand{\eea}{\end{eqnarray}}
\newcommand{\bean}{\begin{eqnarray*}}
\newcommand{\eean}{\end{eqnarray*}}

\renewcommand{\phi}{\varphi}

\renewcommand{\phi}{\varphi}

\preprint{
SAGEX-19-17-E}
\title{Scalar-Graviton Amplitudes} 
\author[a]{N.~E.~J.~Bjerrum-Bohr,}
\author[a]{Andrea~Cristofoli,}
\author[a]{Poul~H.~Damgaard,}
\author[a,b]{Humberto~Gomez.}
\affiliation[a]{Niels Bohr International Academy and Discovery Center\\ 
The Niels Bohr Institute, University of Copenhagen\\
Blegdamsvej 17, DK-2100 Copenhagen, Denmark}
\affiliation[b]{Facultad de Ciencias, Basicas Universidad\\ Santiago de Cali,
Calle 5 $N^\circ$ 62-00 Barrio Pampalinda\\ Cali, Valle, Colombia}
\emailAdd{bjbohr@nbi.dk}
\emailAdd{a.cristofoli@nbi.ku.dk}
\emailAdd{phdamg@nbi.dk}
\emailAdd{humberto.gomez@nbi.ku.dk}

\date{\today}
\abstract{Using the CHY-formalism and its extension to a double cover we provide covariant expressions for tree-level amplitudes with two massive scalar 
legs and an arbitrary number of gravitons in $D$ dimensions. 
Using unitarity methods, such amplitudes are needed inputs for the computation of post-Newtonian and 
post-Minkowskian expansions in classical general relativity. }
\begin{document}
\maketitle
\newpage
%%%%%%%%%%%%%%%%%%%%%%%%%%%%%%%%%%%%%%%%%
\section{Introduction}\label{sec:introduction}
%%%%%%%%%%%%%%%%%%%%%%%%%%%%%%%%%%%%%%%%%
Recently it has been realized that modern methods for amplitude computations at loop level may provide a
powerful new way to compute post-Newtonian and post-Minkowskian expansions in classical general relativity
\cite{Neill:2013wsa,Holstein:2008sx,Bjerrum-Bohr:2013bxa,Vaidya:2014kza,Bjerrum-Bohr:2014zsa,Porto:2016pyg,Damour:2016gwp,Cachazo:2017jef,Damour:2017zjx,Bjerrum-Bohr:2018xdl,Levi:2018nxp,Cheung:2018wkq,Chung:2018kqs,Bern:2019nnu,Antonelli:2019ytb,Cristofoli:2019neg,KoemansCollado:2019ggb,Maybee:2019jus}. This builds on the observation that the quantum mechanical scattering matrix for matter interacting 
gravitationally contains classical pieces at arbitrarily high order in the loop expansion
\cite{Iwasaki:1971vb,Holstein:2004dn,Kosower:2018adc} and the fact that the sought-for long-distance contributions 
are non-analytic in the exchanged momentum \cite{Donoghue:1994dn,BjerrumBohr:2002kt}, thus making them straightforwardly 
accessible through unitarity cuts. All needed contributions being classical, one would not expect it to be
necessary to regularize the loops dimensionally. However, since infrared 'super-classical' (see, \eg,
ref. \cite{Kosower:2018adc}) terms appear at intermediate steps it is nevertheless convenient to
use dimensional regularization. \\[5pt]
For the scattering of two massive objects at large distances the needed tree-level amplitudes are those
of two massive scalars and, at $n$-loop order, $(n+1)$ on-shell gravitons. Using the Kawai-Lewellen-Tye (KLT)
relations \cite{Kawai:1985xq,Bern:1998sv,BjerrumBohr:2010yc,BjerrumBohr:2010hn} these can conveniently be constructed from
the corresponding amplitudes with the $(n+1)$ gravitons replaced by gluons, amplitudes that are given in the
literature on the basis of recursion relations \cite{Badger:2005zh,Forde:2005ue} in
four space-time dimensions, using the spinor-helicity formalism. More recently, Naculich
\cite{Naculich:2014naa} has suggested an alternative and more direct method for the computation 
of such amplitudes based on the Cachazo-He-Yuan (CHY) formalism \cite{Cachazo:2013hca,Cachazo:2014xea}. One
advantage of using the CHY-formalism is that it immediately provides the amplitudes 'covariantly', in terms
of general polarization tensors for the gravitons, and hence not restricted to four space-time dimensions.\\[5pt]
From a practical point of view, it suffices to evaluate amplitudes with two scalar legs and $(n+1)$ gluons
and subsequently turning them into scalar-graviton amplitudes by KLT-squaring. This is our approach here.
One key point of the present calculation is the computation of a factorized expression for the amplitudes of two
massive scalars coupled to Yang-Mills theory, expressing them as sums over lower-point amplitudes which
are combinations of scalar-gluon amplitudes and pure gluon amplitudes. Each of these has one gluon leg
off-shell and an associated polarization vector of both transverse and longitudinal components. 
In this way, we can iteratively construct amplitudes of an arbitrarily high order. Crucial for this factorized form 
is the insight gained from the double-cover version 
\cite{Gomez:2016bmv,Cardona:2016bpi,Bjerrum-Bohr:2018lpz,Bjerrum-Bohr:2018jqe,Gomez:2018cqg} of the CHY-formalism. This double-cover
description naturally splits amplitudes into two lower-point amplitudes, each with one leg off-shell. These vector
currents, contracted with polarization vectors, are glued together by the polarization sum. A subtlety here is the
contribution from longitudinal modes that need to be dealt with carefully. Useful relations that short-cut the evaluations
of some of the color-ordered amplitudes needed for the recursive evaluation of higher $n$-point amplitudes are provided by
simple identities \cite{Bjerrum-Bohr:2016juj,Cardona:2016gon,Bjerrum-Bohr:2016axv} among these partly massive amplitudes. \\[5pt]
The outline of this paper is as follows. In sections 2 and 3 we show how to compute amplitudes with two scalars and $n$ gluons using different methods. In section 4 we briefly discuss the straightforward application of Kawai-Lewellen-Tye relations to replace the gluons with gravitons. Some technical details and a proof of an important theorem regarding vanishing longitudinal contributions are provided in appendices.
%%%%%%%%%%%%%%%%%%%%%%
\section{Prelude: Two massive scalars and  $n$ gluons}\label{SGluons}
%%%%%%%%%%%%%%%%%%%%%%
We first present a simple way to obtain explicit expressions for the scattering amplitudes of two massive scalars and $n$ gluons. Since our method is based on the CHY approach, we give a very brief review of this formalism. We then apply the factorization method developed in \cite{Gomez:2018cqg} to obtain, up to six-point, analytical expressions for the scattering of gluons where two of them, suitably defined, are massive. Next, we turn the two massive gluons into massive scalars, thus providing the scattering amplitudes for two massive scalars and in principle any number of massless gluons. 
\subsection{Massive Yang-Mills Amplitudes}\label{Off-shell-YM}
We start by presenting a simple recursive formula that computes pure Yang-Mills amplitudes with up to three massive gluons. The method we will use was developed by one of us in a different context \cite{Gomez:2016bmv,Gomez:2018cqg}. We shall show explicit expressions up
to six points but it is straightforward to extend the method to any higher number of external legs. In the following, we will denote massive particles with the capital letter ``$P_\a$" and the massless ones with the lower-case letter ``$k_a$". Unless otherwise mentioned we will work under the assumption of implicit momentum conservation,
\begin{equation}
K_1+K_2+\cdots + K_n = 0\,.
\end{equation}
Let us first recall how to extend the CHY approach to the massive case following the method of Naculich \cite{Naculich:2014naa}. We have $\{ P_1,...,P_i \}$ as momenta of the massive particles ($P_\a^2\neq 0$) and  $\{ k_{i+1},...,k_n \}$ as momenta the massless gluons ($k_a^2=0$). A generic momentum vector is thus $K_A \in \{ P_1,...,P_i, k_{i+1},...,k_n \}$. We define as well 
\begin{eqnarray}
&& P_{AB\ldots D} \equiv K_A+K_B+\cdots + K_D, \nonumber\\
&& P_{A:A+j} \equiv K_A+K_{A+1}+\cdots + K_{A+j}\,.
\end{eqnarray} 
The modified CHY scattering equations are then given by
%%%%%%%%%%%%%%%%%%%%%%%
\begin{equation}
S_{A}=\sum_{B=1 \atop B\neq A}^n \frac{2\, K_A \cdot K_B + 2\, \Delta_{AB} }{\s_{A B}} =0\,, \quad A=1,2,...,n\,,
\end{equation}
%%%%%%%%%%%%%%%%%%%%%%%%%
where the matrix $\Delta_{AB}$ is still to be determined.
In order to guarantee ${\rm SL}(2,\mathbb{C})$ invariance, \ie, $\sum_{A=1}^n \s_A^m S_A =0$ for $m=0,1,2$, the matrix $\Delta_{AB}$ must be symmetric, $\Delta_{AB}=\Delta_{BA}$, and it must satisfy the conditions
%%%%%%%%%%%%%%%%%%%%%%%%%
\begin{eqnarray}
&& \sum_{\b=1 \atop \b \neq a}^i \Delta_{\a\b} + \sum_{b = i+1}^n \Delta_{\a b} = P_\a^2\,, \quad \a=1,\ldots , i, \nonumber\\
&& \sum_{\b=1 }^i \Delta_{a\b} +\sum_{b = i+1 \atop b\neq a}^n  \Delta_{ab} =0\,, \quad \quad a=i+1,\ldots , n .
\end{eqnarray}
%%%%%%%%%%%%%%%%%%%%%%%%%
Since we are interested in at most up to three massive gluons of momenta $\{P_1,P_2,P_3\}$, it is sufficient to consider only $\Delta_{12}, \Delta_{13}, \Delta_{23}$. Therefore, we therefore have the simple conditions
%%%%%%%%%%%%%%%%%%%%%%%
\begin{eqnarray}
&& \Delta_{12}+\Delta_{13} = P_1^2\,,  \nonumber \\
&& \Delta_{12}+\Delta_{23} = P_2^2\,, \\
&& \Delta_{13}+\Delta_{23} = P_3^2 \,, \nonumber
\end{eqnarray}
%%%%%%%%%%%%%%%%%%%%%%%%
that have a unique solution given by
\begin{eqnarray}\label{solthree}
\Delta_{12}=\frac{ P_1^2 +P_2^2 - P_3^4}{2}\,, \quad  \Delta_{13}= \frac{ P_1^2 -P_2^2 + P_3^4}{2}\,, \quad
 \Delta_{23}= \frac{-P_1^2 +P_2^2 + P_3^4}{2}\,. \qquad
\end{eqnarray}
%%%%%%%%%%%%%%%%%%%%%%%%%%%
When two masses are degenerate, \eg\,, $P_1^2=P_2^2 \neq 0$ and $P_3^2=0$, it is straightforward to see from \eqref{solthree} that $\Delta_{12}=P_1^2$ and $\Delta_{13}=\Delta_{23}=0$, which, not surprisingly, is in agreement with the one-loop scattering equations formulated in refs. \cite{Adamo:2013tsa,Casali:2014hfa,Geyer:2015jch,Cardona:2016bpi}. On the other hand, when only one of the legs is massive, \eg\, $P_1^2 \neq 0$ and $P_2^2 = P_3^2=0$, then $\Delta_{12}=P_1^2/2$, $\Delta_{13}=P_1^2/2$ and $\Delta_{23}=-P_1^2/2$, \ie, in order to describe one massive particle 
it is necessary to use at least three $\Delta_{AB}$ parameters.\\[5pt]
After having described the massive scattering equations let us now remind that the CHY prescription for color ordered amplitudes of the scattering of gluons at tree-level is given by \cite{Cachazo:2013hca,Cachazo:2013iea,Naculich:2014naa}
\be\label{chyYM}
A_n(P_1,...,P_i, i+1, ... ,n)= \int d\mu_n \, 
{\rm PT}{(1,2,...,n)}
\times {\rm Pf}^\prime \Psi_n\,, \ee
where $d\mu_n$ is the usual CHY measure
\begin{equation}\label{measure}
d\mu_n = (\sigma_{jk} \sigma_{kl} \sigma_{lj}) \!\! \prod^n_{A=1 \atop A\neq j,k,l} \!\! \! d\sigma_A 
\times
 (\sigma_{mr} \sigma_{rs} \sigma_{sm})\!\!
 \prod^n_{B=1 \atop B\neq m,r,s} \!\!\! \delta(S_B)\,,
\end{equation}
%%%%%%%%%%%%%%%%%%%%%%%%%%%%%
and ${\rm PT}{(1,...,n)}$ and $ {\rm Pf}^\prime \Psi_n$ are the usual Parke-Taylor and reduced Pfaffian factors  
\be\label{PTandRP}
{\rm PT}{(1,...,n)}\equiv \frac{1}{\sigma_{12} \sigma_{23} \cdots \sigma_{n1}}\,, \quad
{\rm Pf}^\prime \Psi_n \equiv \frac{(-1)^{A+B}}{ \sigma_{AB} }\,\,{\rm Pf}[(\Psi_n)^{AB}_{AB}]\,.
\ee
The $2n\times 2n$ matrix, $\Psi_n$, is defined as 
\vspace{-0.0cm}
\begin{eqnarray}\label{Pmatrix}
\Psi_n \equiv \left( 
\begin{matrix}
\mathsf{A} & -\mathsf{C}^{\rm T} \\
\mathsf{C} & \mathsf{B} 
\end{matrix}
\right)\,, 
\end{eqnarray}\vskip-0.1cm\noindent
with, 
\be
\mathsf{A}_{AB} \equiv 
\begin{cases} \displaystyle 
\frac{ 2 \, K_{A}\cdot K_B + 2\, \Delta_{AB} }{ \sigma_{AB} }\,, \\
\displaystyle \quad ~~ 0\,, \end{cases} \!\!\!\!\!\!\! \ \ \ \ \ 
\mathsf{B}_{AB} \equiv 
\begin{cases} \displaystyle \frac{\eps_A\cdot \eps_A}{\sigma_{AB}} & A\neq B\,,\\
\displaystyle \quad ~~ 0 & A=B\,,
\end{cases}
\ee
and\vskip-0.7cm
\be\label{C-matrix}
\mathsf{C}_{AB} \equiv  \begin{cases} \displaystyle \frac{\sqrt{2} \, \eps_{A}\cdot K_B}{\sigma_{AB} }\,, & A\neq B\,,\\
\displaystyle -\sum_{C=1 \atop C\neq A}^n \frac{\sqrt{2} \, \epsilon_A \cdot K_C}{ \sigma_{AC} }\,, & A=B\,.
\end{cases}
\ee\vskip-0.0cm\noindent
The matrix, $(\Psi_n)^{AB}_{AB}$, denotes the reduced matrix obtained by removing the rows and columns $A, B$ from $\Psi_n$, where $1 \leq A< B \leq n$. \\[5pt]
Since we are interested in the case of at most three massive particles of momenta $\{P_1,P_2,P_3 \}$ we can avoid dealing with the $\Delta_{AB}$-matrix in the scattering equations altogether by choosing the labels $\{ j,k,l\}$ and $\{ m,r,s \}$ in \eqref{measure} to match with the massive ones, \ie, $\{ j,k,l\}=\{ m,r,s \}=\{ 1,2,3 \}$. \\[5pt]
It is useful to recall that the reduced Pfaffian ($ {\rm Pf}^\prime \Psi_n=\frac{(-1)^{A+B}}{ \sigma_{AB} }\,\,{\rm Pf}[(\Psi_n)^{AB}_{AB}]$) is independent of the choice of $A$ and $B$, and that the ${\rm SL}(2,\mathbb{C})$ symmetry is guaranteed by the transversality of the external polarization vectors, $(\epsilon_C \cdot K_C )=0$. However, we note that the terms $C_{AA}$ and $C_{BB}$ do not appears in the reduced matrix, $(\Psi_n)^{AB}_{AB}$. It follows that the transversality conditions on $\epsilon_A$ and $\epsilon_B$ are not needed to obtain an integrand invariant under the action of ${\rm SL}(2,\mathbb{C})$ \cite{Dolan:2013isa}. We can therefore consistently define the integral with these two legs being off mass-shell and with arbitrary polarization vectors for $(\epsilon_A \cdot K_A )\neq 0$ and $(\epsilon_B \cdot K_B )\neq 0$. 
We now use the double-cover method ref. \cite{Gomez:2018cqg} to obtain compact recursive expressions for these massive and/or off-shell scattering amplitudes as defined above. The results clearly reduce to the usual expressions when all external legs are massless and on-shell. \\[5pt]
%%%%%%%
First, let us consider the basic building block of three legs. We take all three particles to be massive and choose the polarization vectors $\eps_1$ and $\eps_2$ as not necessarily transverse so that we do not impose $(\eps_1\cdot { P_1}) = 0 = (\eps_2\cdot { P_2})$. We are going to denote with a $\bm{bold}$ source in the amplitude (as in. \cite{Cachazo:2013hca,Naculich:2014naa,Gomez:2018cqg}), \eg 
\be
A_n(\ldots , {\bm P}_{\a}, \ldots , {\bm P}_{\b}, \ldots )\,,
\ee
the rows/columns that are removed from its reduced Pfaffian. In the above amplitude the reduced Pfaffian is given by, $ {\rm Pf}^\prime \Psi_n=\frac{(-1)^{\a+\b}}{ \sigma_{\a\b} }\,\,{\rm Pf}[(\Psi_n)^{\a\b}_{\a\b}]$. Particles $P_\a$ and $P_\b$ can thus be off-shell, so that $(\epsilon_\a\cdot P_\a) \neq0$ and $(\epsilon_\b\cdot P_\b )\neq0$.\\
Therefore, using the CHY prescription given in \eqref{chyYM} one has
%%%%%%%%%%%%%%%%%%%%%%%%%%%%
\begin{eqnarray}\label{YM123}
A_3 ({\bm P_1},{\bm P_2}, {P_3} )  
\! &=&\!
(\s_{12}\,\s_{23}\,\s_{23})^2 \, {\rm PT}(1,2,3)
\,
\frac{(-1)}{\s_{12}}
\,
{\rm Pf}\left[
{\small
\begin{matrix}
0 & - \frac{\eps_{1}\cdot \sqrt{2} \, P_{3}}{\s_{13}} & - \frac{\eps_{2}\cdot \sqrt{2}\, P_{3}}{\s_{23}} & - \mathsf{C}_{33}\\
 \frac{\eps_{1 }\cdot \sqrt{2}\, P_{3}}{\s_{13}} & 0 & \frac{\eps_{1}\cdot \eps_{2}}{\s_{12}} & \frac{\eps_{1}\cdot \eps_{3}}{\s_{13}}\\
 \frac{\eps_{2}\cdot \sqrt{2} \,P_{3}}{\s_{23}} & \frac{\eps_{2}\cdot \eps_{1}}{\s_{21}} & 0 & \frac{\eps_{2}\cdot \eps_{3}}{\s_{23}} \\
\mathsf{C}_{33} &  \frac{\eps_{3}\cdot \eps_{1}}{\s_{31}} & \frac{\eps_{3}\cdot \eps_{2}}{\s_{32}} & 0 \\
\end{matrix}}
\right] 
\nonumber \\
& =& 
\sqrt{2} \left\{
(\eps_1\cdot \eps_2)(\eps_3\cdot { P_1 }) - (\eps_2\cdot \eps_3) (\eps_1\cdot { P_3}) +(\eps_3\cdot \eps_1)(\eps_2\cdot { P_3})
\right\}\,, \qquad 
\end{eqnarray}
%%%%%%%%%%%%%%%%%%%%%%%%%
where we have used 
\be
\mathsf{C}_{33}=-\sqrt{2} \,\left( \frac{\eps_{3}\cdot P_{1}}{\s_{31}} +\frac{\eps_{3}\cdot P_{2}}{\s_{32}} \right) = \sqrt{2} \, (\eps_3\cdot P_1) \times \frac{ \s_{12}}{ \s_{31} \, \s_{23} }\,,
\ee 
due to the momentum conservation constraint ${P_1}+{ P_2}+{ P_3}=0$ and the transversality condition $(\eps_3\cdot { P_3}) = 0 $. Although the amplitude itself is independent of the choice of rows/columns that are removed in the Pfaffian, the intermediate expressions do depend on the choice and we have therefore introduced a notation where we indicate which rows and columns are removed.\\[5pt]
%%%%%%%%%%%%%%%%%%%%%%%%%%%%%%%%%%%%%%%%%%%%%
We consider next a computation with three massive gluons of momenta $\{{ P_1},{ P_2},{ P_3} \}$ and one massless gluon of momentum $\{k_4\}$. Using the factorization method described in \cite{Gomez:2018cqg,Bjerrum-Bohr:2018lpz}, this four-point calculation can be expressed in terms of the $A_3({\bm P_a},{\bm P_b},{ P_c})$ building-blocks,  
%%%%%%%%%%%%%%%%%%%%%%%%%%%%%%
%
\begin{eqnarray}\label{YM4points}
&&A_4 ({\bm P_1},P_2,{\bm P_3},4 ) \nonumber\\
&&=\sum_M\left [ \frac{ A_3 ({\bm P^{\eps^M}_{34}},{\bm P_1}, {P_2} ) \, A_3 ({\bm P^{\eps^M}_{12}},{\bm P_3}, 4 )   }{s_{P_34}} + \frac{ A_3 ({\bm P_1},{\bm P^{\eps^M}_{23}}, {4} ) \, A_3 ({\bm P_3},{\bm P^{\eps^M}_{41}}, {P_2} )   }{s_{4P_1}} \right] \nonumber \qquad \qquad\\
&&\hskip5.7cm- 2 \sum_L\left[ \frac{  A_3 ({\bm P^{\eps^L}_{13}},{\bm P_2}, {4} )}{s_{P_24}} \times A_3 ({\bm P^{\eps^L}_{24}},{\bm P_1}, {P_3} ) \right]\,,
 \qquad 
\end{eqnarray}
%
%%%%%%%%%%%%%%%%%%%%%%%%%%%%%%
where the notation $P^{\eps^M}_{i}$ ($P^{\eps^L}_{i}$) means the particle with momentum $P_i$ has as polarization vector $\eps_i^{M}$ ($\eps_i^{L}$). The sums over the polarizations are given by the relations
\begin{eqnarray}
&& \sum_M \eps^{M\, \mu}_i \eps^{M\, \nu}_j = \eta^{\mu\nu}\,,  \label{Eta}\\
&& \sum_L \eps^{L\, \mu}_{i} \eps^{L\, \nu}_{j} = \frac{P^\mu_{i}\, P^\nu_{j} }{P_{i}\cdot P_{j} + P_1^2 - P_3^2}\,. \label{Longitudinal}
\end{eqnarray}
The unusual normalization factor of the longitudinal modes is precisely what is needed to recover the correct four-point amplitude \cite{Bjerrum-Bohr:2018jqe,Gomez:2019cik}.
The polarization vectors of all massive on-shell legs of course still satisfy $\epsilon_i\cdot P_i = 0$. Using that 
condition it is easy to see that the last term in \eqref{YM4points} evaluates to
%%%%%%%%%%%%%%%%%%%%%%%%%%%%
\begin{eqnarray}\label{Lcontribution}
- 2 \sum_L\left[\frac{  A_3 ({\bm P^{\eps^L}_{13}},{\bm P_2}, {4} )}{s_{P_24}} \times A_3 ({\bm P^{\eps^L}_{24}},{\bm P_1}, {P_3} ) \right] 
= (\eps_1\cdot \eps_3)(\eps_2\cdot \eps_4)\,.
 \qquad 
\end{eqnarray}
The full four-point amplitude is thus remarkably simple.\\[5pt]
%%%%%%%%%%%%%%%%%%%%%%%%%%%  
Finally, in order to calculate higher-point amplitudes we will also need $A_4 ({\bm P_1},{\bm P_2},P_3,4 ) $. Using the BCJ-like identity \cite{Bjerrum-Bohr:2016juj,Cardona:2016gon,Bjerrum-Bohr:2016axv}, 
$$
s_{4P_3} \, {\rm PT}(3,4,1,2) + s_{4P_{13}} \, {\rm PT}(3,1,4,2)=0\,,
$$
it is straightforward to deduce
%%%%%%%%%%%%%%%%%%%%%%%%%%%%%%
\begin{eqnarray}\label{YM4pts2}
A_4 ({\bm P_1},{\bm P_2},P_3,4 ) 
=-
\left( 1+ \frac{s_{4 P_1}}{s_{4 P_3}} \right) \times
A_4 ({\bm P_1},P_3,{\bm P_2},4 )\,.
\end{eqnarray}
%%%%%%%%%%%%%%%%%%%%%%%%%%%%%%
The calculation of higher-point amplitudes with massive gluons now proceeds recursively. We illustrate a few cases in the appendix.
\subsection{Turning massive gluons into scalars}
Now, using the prescriptions of Naculich \cite{Naculich:2014naa} and Cachazo, He, and Yuan \cite{Cachazo:2014xea} we can compute the amplitudes of interest which also involve massive scalar legs.
The basic idea is to consider the massive gluon theory in one extra dimension (\ie, in $D+1$ dimensions) with ``polarizations'' and momenta of massive scalars chosen to be
%%%%%%%%%%%%%%%%%%%%%%%%%
%
\begin{eqnarray}
&&
\left.
\begin{matrix}
P^\mu_1=(\vec{p_1},0), \quad \eps_1^\mu = (\vec{0},1) \, \, \, \, \, \\
P^\mu_n=(\vec{p_n},0), \quad \eps_n^\mu = (\vec{0},1) \, \, \, \, \, \\
\end{matrix}
\right\} \quad \text{Massive scalars }\, (P_1^2=P_n^2 = m^2)\,,  \\
&&
\left.
\begin{matrix}
k^\mu_a=(\vec{k_a},0), \quad \eps_a^\mu = (\vec{\eps}_a,0) \, \, \, \, \, \\
\end{matrix}
\right\} \quad \text{Massless gluons }\, (a=2,...,n-2, \,\, {\rm and} \, \, k_a^2=0)\,. \qquad
\nonumber
\end{eqnarray}
%
%%%%%%%%%%%%%%%%%%%%%%%%%
In this set-up all external particles satisfy $P_i\cdot\epsilon_i = 0 =k_a\cdot\epsilon_a$ and, additionally, it is easy to see that 
$\Delta_{1n}=\Delta_{n1}=m^2$
 in Naculich's notation as a consequence of equation \eqref{solthree}. 
 %%%%%%%%%%%%%%%%%%%%%%%%%%%%%
The CHY prescription for the scattering of two massive scalars with $n-2$ gluons can thus be written as
%%%%%%%%%%%%%%%%%%%%%%%%
\begin{eqnarray}\label{AnSgluons}
A _n (1_\phi, 2_g, ..., (n-1)_g,n_\phi) = \int d\mu_n \,\, {\rm PT}(1,2,...,n)  \times {\rm Pf}^\prime \Psi_n\Big|_{\substack{ \eps_1,\eps_n=(\vec{0},1) }} \, ,
\end{eqnarray}
%%%%%%%%%%%%%%%%%%%%%%%%
where the massive scattering equations, the reduced Pfaffian and the measure as defined above.
It is useful to note that these ordered amplitudes are invariant under cyclic permutations, \ie,
\begin{eqnarray}\label{cyc}
A_n(1,2,... ,{\bm P_\a},...,{\bm P_\b},...,n ) = A_n(n,1,2,... ,{\bm P_\a},...,{\bm P_\b},...,n-1 )\,, 
\qquad
\end{eqnarray}
and also satisfy
\begin{eqnarray}
A_n(1,2,... ,{\bm P_\a},...,{\bm P_\b},...,n-1,n ) =(-1)^n A_n(n,n-1,... ,{\bm P_\b},...,{\bm P_\a},...,2,1 )\,. \nonumber
\end{eqnarray}
As a first step we note that when $P_{1}$ and $P_{2}$ are associated with scalar legs the three-point amplitude reads
\begin{eqnarray}
A_3 ({\bm P_1},{\bm P_2}, {P_3} )\Big|_{\eps_1,\eps_2=(\vec{0},1)} = \sqrt{2}\, (\eps_3\cdot { P_1})\,.
\end{eqnarray}
We illustrate our method by evaluating the four-point function of two massive scalars and two gluons. Using the above conditions and the cyclicity property \eqref{cyc} we immediately infer this amplitude from eqs. \eqref{YM4points}:
%%%%%%%%%%%%%%%%%%%%%%%%
\begin{eqnarray}\label{A4pts}
&&
A_4(1_\phi, 2_g, 3_g, 4_\phi) 
= A_4({\bm P_4}, P_1, {\bm 2}, 3)\Big|_{\substack{ \eps_1,\eps_4=(\vec{0},1) }} \\
&&=\sum_M\left [ \frac{ A_3 ({\bm P^{\eps^M}_{23}},{\bm P^\phi_4}, {P^\phi_1} ) \, A_3 ({\bm P^{\eps^M}_{41}},{\bm 2}, 3 )   }{s_{23}} + \frac{ A_3 ({\bm P^\phi_4},{\bm P^{\eps^M}_{12}}, {3} ) \, A_3 ({\bm 2},{\bm P^{\eps^M}_{34}}, {P^\phi_1} )   }{s_{3P_4}} \right],
\nonumber
\end{eqnarray}
%%%%%%%%%%%%%%%%%%%
where the superscript (or subscript) ``$\phi$" refers to one of the massive scalars. We note that the term 
\eqref{Lcontribution} does not contribute at all, (we shall return to this point later).\\[5pt]
{\bf Remark:}
Since $\eps^\mu_1=\eps^\mu_4=(\vec{0},1)$ the contraction relation for the second term in \eqref{A4pts}, 
\begin{equation}
\sum_{M} (\eps_{P_{34}}^{M} \cdot \eps_1) (\eps_{P_{12}}^{M}\cdot V)= (\eps_1\cdot V)\,,
\end{equation}
is non-vanishing only when $V^\mu$ has a non-zero projection on $\epsilon_4$. Therefore, it is equivalent to choosing $\eps_{P_{34}}^{M\, \mu}=\eps_{P_{12}}^{M\, \mu} = (\vec{0},1)$, \ie, the internal lines corresponding to momenta ${P_{12}}$ and ${P_{34}}$ turn out to be propagating scalars as expected due to current conservation. In other words,
%%%%%%%%%%%%%%%%
\begin{equation}\label{Mtoscalar}
\sum_M \frac{ A_3 ({\bm P^\phi_4},{\bm P^{\eps^M}_{12}}, {3} ) \, A_3 ({\bm 2},{\bm P^{\eps^M}_{34}}, {P^\phi_1} )   }{s_{3P_4}} =
 \frac{ A_3 ({\bm P^\phi_4},{\bm P^{\phi}_{12}}, {3} ) \, A_3 ({\bm 2},{\bm P^{\phi}_{34}}, {P^\phi_1} )   }{s_{3P_4}}\,. 
\end{equation}
%%%%%%%%%%%%%%%%
The same phenomenon occurs for higher $n$-point amplitudes.
Let us now introduce some convenient notation:
\begin{eqnarray}
&& F^{\mu\nu}_{A}\, \equiv\, K^\mu_A\,\eps^\nu_A-K^\nu_A\, \eps^\mu_A\, , \nonumber\\
&& (V\,W)_{ab}\, \equiv\, V_a^\mu \, \eta_{\mu\nu}\, W_b^\nu\,,
\end{eqnarray}
as well as
\begin{eqnarray}
&& (V \, F \ldots F \, W)_{aA_1\ldots A_j b}\, \equiv\,  V_a^\mu \, \eta_{\mu \gamma} \, F_{A_1}^{\gamma\nu} \, \eta_{\nu\sigma} \, F_{A_2}^{\sigma\alpha} \, \cdots \, F_{A_j}^{\rho\delta}\,  \eta_{\delta \b}  W_b^{\b} \,, \\
&& s_{A_1A_2\ldots A_j}\, \equiv\, (K_{A_1}+K_{A_2}+\cdots + K_{A_j})^2 - (K^2_{A_1}+K^2_{A_2}+\cdots + K^2_{A_j})\,, \nonumber \qquad
\end{eqnarray} 
where $V_a^{\mu}$ and $W_b^\nu$ are two generic vectors.
From  \eqref{Eta} and \eqref{YM123} it is straightforward to compute
%%%%%%%%%%%%%%%%%%%%%%%%
\begin{eqnarray}\label{cut-1-4pts}
\sum_M \frac{ A_3 ({\bm P^{\eps^M}_{23}},{\bm P^\phi_4}, P_1^\phi ) \, A_3 ({\bm P^{\eps^M}_{41}},{\bm 2}, 3 )   }{s_{23}} 
=\frac{ 2\, (\eps P)_{21} \, (\eps k)_{32} - 2\, (\eps F P )_{231} }{ s_{23} }\,,
\end{eqnarray}
%%%%%%%%%%%%%%%%%%%
as well as
%%%%%%%%%%%%%%%%%%%%%%%%
\begin{eqnarray}\label{cut-2-4pts}
 \frac{ A_3 ({\bm P^\phi_4},{\bm P^{\phi}_{12}}, {3} ) \, A_3 ({\bm 2},{\bm P^{\phi}_{34}}, {P^\phi_1} )   }{s_{3P_4}} 
=- \frac{2 (\eps P)_{21} \, (\eps P)_{34} }{s_{P_12}}\,.
\end{eqnarray}
%%%%%%%%%%%%%%%%%%%
The four-point covariant amplitude of two massive scalars and two gluons is thus given by the simple expression
%%%%%%%%%%%%%%%%%%%%%%%%
\begin{eqnarray}\label{A4pts2}
A_4(1_\phi, 2_g, 3_g, 4_\phi) 
=
\frac{ 2\, (\eps P)_{21} \, (\eps k)_{32} - 2\, (\eps F P )_{231} }{ s_{23} }
- \frac{2 (\eps P)_{21} \, (\eps P)_{34} }{s_{P_12}}\,.
\qquad\quad
\end{eqnarray}
%%%%%%%%%%%%%%%%%%%
Specializing to four dimensions, this is in agreement with the result found in the literature on the basis of the spinor-helicity formalism \cite{Forde:2005ue}.\\[5pt]
%%%%%%%%%%%%%%%%%%%%%%%%%%%
In an analogous way, the five-point amplitude becomes
%%%%%%%%%%%%%%%%%%%%%%%%
\begin{eqnarray}\label{A5pts}
&&
A_5(1_\phi, 2_g, 3_g, 4_g,5_\phi) 
= A_5({\bm P_5}, P_1, {\bm 2}, 3,4)\Big|_{\substack{ \eps_1, \eps_5=(\vec{0},1) }} \nonumber \\
&&
=(-1)\times \sum_M\left[ \frac{ A_3 ({\bm P^{\eps^M}_{4:1}},{\bm 2}, {3} ) \, A_4 ({\bm P^\phi_5}, P^\phi_1, {\bm P^{\eps^M}_{23}}, 4)}{s_{23}} +  
 \frac{ A_3 ({\bm P^{\eps^M}_{2:4}},{\bm P^\phi_{5}}, {P^\phi_1} ) \, A_4 ({\bm P^{\eps^M}_{51}}, {\bm 2}, 3, 4 )   }{s_{234} }
\right]
\nonumber\\
&&
\hspace{2.0cm}
 +
 (-1)\times 
 \frac{ A_3 ({\bm {2}},{\bm P^{\phi}_{3:5}}, {P^\phi_1} ) \times A_4 ({\bm P^\phi_5}, {\bm P^{\phi}_{12}} , 3, 4 )   }{s_{34P_5} }\,, 
\end{eqnarray}
%%%%%%%%%%%%%%%%%%%
where eq. \eqref{YM5pts} has been used. As in the four-point case, the purely longitudinal contributions vanish on account of the orthogonality conditions for the polarization vectors associated with external scalar legs, $(\eps_1\cdot \eps_3)= (\eps_5\cdot \eps_2) =0$. In appendix \ref{preposition}, we prove the vanishing of these longitudinal contributions  
for any number of external gluons.\\[5pt]
Applying the identity \eqref{Eta}  and using \eqref{YM123}, \eqref{YM4points} and \eqref{YM4pts2} we finally find an explicit covariant expression for $A_5(1_\phi, 2_g, 3_g, 4_g,5_\phi) $:
%%%%%%%%%%%%%%%%%%%%%%%%
\begin{eqnarray}\label{A5expression}
&&
\left( \frac{1}{\sqrt{2} } \right)\times A_5(1_\phi, 2_g, 3_g, 4_g,5_\phi) =
(\eps P)_{21}\times 
\frac{  
( \eps \eps)_{ 34 } \,  s_{ 34 } 
-
2\, (\eps k)_{34} \, (\eps P)_{ 45 } + 
 2\, (\eps k)_{ 43 } \, ( \eps P)_{ 35 }   
} {s_{P_12} \, s_{34}} 
\nonumber \\
&&
+
\frac{
 (\eps \eps)_{23}\, ( \eps k)_{ 43 } s_{ 23 }  
 - (\eps \eps)_{34}\, ( \eps P)_{ 21 } s_{ 23 }   
-  
 ( \eps \eps)_{ 23 } \, ( \eps P )_{ 41 }\, s_{ 34 }  
-
 (\eps F \eps)_{342} \, s_{ 23 }  
 }
 {s_{23} \, s_{34}} +
  \frac{
 ( \eps \eps)_{ 34 } \, (\eps P)_{ 21 } \, s_{4P_5}
 }
 {s_{P_12} s_{34}} 
\nonumber \\
&&
+
2(\eps P)_{21} ( \eps P)_{ 45 } \times 
\frac{
 ( \eps P)_{ 31 } + 
  ( \eps k)_{ 32 }  
 }
 {s_{P_1 2} \, s_{4P_5}} +
 (\eps P)_{ 45 } \times 
\frac{
2  ( \eps k)_{ 21 }   ( \eps P)_{ 32 } 
-
2 ( \eps k)_{ 23 }  ( \eps P)_{ 31 } -
 ( \eps \eps)_{ 23 }  s_{23}
 }
 {s_{ 23} \, s_{4P_5}} 
 \nonumber  \\ 
&&  
+\frac{
 s_{P_12 }  (\eps F \eps)_{ 243 }   
 +
  (\eps \eps)_{ 23 } \,  (\eps k)_{ 43 } \,s_{ P_12 } 
  - 
  (\eps \eps)_{ 34 } \,  (\eps P)_{ 21 } \,s_{ P_14 } - 
  (\eps \eps)_{ 34 } \,  (\eps P)_{ 25 } \,s_{ P_14 } + 
  (\eps \eps)_{ 34 } \,  (\eps P)_{ 21 } \,s_{ 23 }}
  {s_{34} \, s_{234}} 
    \nonumber\\   
 &&
 +
 \frac{
 (\eps \eps)_{ 23 } (\eps P)_{ 41 } s_{34}   
 }
 {s_{23} s_{234} } 
 -
 \frac{
 ( \eps \eps)_{ 23 } \, (\eps P)_{ 45 } \, s_{P_12}
 }
 {s_{23} s_{4P_5}} 
 +
\frac{
(\eps \eps)_{34 } (\eps P)_{ 21 } 
- (\eps \eps)_{ 24 } (\eps P)_{ 31 } 
+ ( \eps \eps)_{ 23 } (\eps P)_{ 41 } 
}
{ s_{234}}
 \nonumber\\
&&
+
\frac{
 (\eps \eps)_{ 34 } (\eps k)_{ 23 } s_{P_14} 
-
 (\eps \eps)_{ 24 } (\eps k)_{ 32 } s_{P_14}  
 +
 (\eps \eps)_{ 23 } (\eps k)_{ 42 } s_{P_14} 
 - 
 (\eps \eps)_{ 23 } (\eps P)_{ 41 } s_{P_12}
 - 
 (\eps \eps)_{ 23 } (\eps P)_{ 45 } s_{P_12}    
 }
 {s_{23} s_{234} } 
+ 
 \nonumber\\  
 &&
2\! \times\!\!
\frac{
 (\eps P)_{ 21 } (\eps k)_{ 32 } (\eps P)_{ 45 } \!
 + \!
  (\eps P)_{ 25 } (\eps P)_{ 31 } (\eps P)_{ 45 } \! 
+ \!
  (\eps P)_{ 25 } (\eps P)_{ 31 } (\eps k)_{ 42 }\!
  +\!
  (\eps P)_{ 21 } (\eps P)_{ 31 } (\eps P)_{ 45 }  \!
  -\!
  (1\!\! \leftrightarrow \! \! 5) }
{s_{34} \, s_{234}}  
\nonumber\\
&&
 + 
2\times 
\frac{
 (\eps k)_{ 23 } (\eps P)_{ 35 } (\eps P)_{ 41 }  
 + 
(\eps k)_{ 32 } (\eps P)_{ 45 } (\eps P)_{ 21 }  
-
(1  \leftrightarrow  5)
 }
{s_{23} s_{234} } \,.
  \qquad
  \nonumber\\ 
\end{eqnarray}
%%%%%%%%%%%%%%%%%%%
Specializing to four dimensions, this matches
the spinor-helicity result provided in \cite{Forde:2005ue}. 
We note that this five-point amplitude $A_5(1_\phi, 2_g, 3_g, 4_g,5_\phi) $ can also be computed using eq.
 \eqref{YM5pts2} so that, alternatively, 
%%%%%%%%%%%%%%%%%%%%
\begin{eqnarray}\label{A5pts2}
&&
A_5(1_\phi, 2_g, 3_g, 4_g,5_\phi) 
= A_5({\bm 3},4, {\bm P_5}, P_1, 2)\Big|_{\substack{ \eps_1, \eps_5=(\vec{0},1) }}\,.
\end{eqnarray}
%%%%%%%%%%%%%%%%%%%%%%%%%%%
It is now straightforward to move to any higher number of points, recursively. Using the result of the appendix we find 
the six-point amplitude  
%%%%%%%%%%%%%%%%%%%%%%%%%%%%%%
\begin{eqnarray}\label{A6pts}
&&
A_6(1_\phi, 2_g, 3_g, 4_g,5_g,6_\phi) 
=
A_6 ({\bm P_6},P_1,{\bm 2},3,4,5 )\Big|_{\substack{ \eps_1, \eps_6=(\vec{0},1)  }} = \\
&&
 \frac{ A_3 ({\bm {2}} , {\bm P^{\phi}_{3:6}}, {P^\phi_1} ) \,  A_5 ({\bm P^\phi_6},{\bm P^{\phi}_{12}},3, 4,5)  }{s_{3 4 5 P_6} }
+
\sum_M\left[  
 \frac{ A_3 ({\bm P^{\eps^M}_{2:5}},{\bm P^\phi_{6}}, {P^\phi_1} ) \, A_5 ({\bm P^{\eps^M}_{61}},{\bm 2},3, 4,5 ) }{s_{2 3 4 5} } 
\right.
\qquad
\nonumber\\
&&
\left.
 +
 \frac{ A_3 ({\bm P^{\eps^M}_{4:1}},{\bm 2}, {3} ) \, A_5 ({\bm P^\phi_6}, P^\phi_1, {\bm P^{\eps^M}_{23}}, 4,5)}{s_{23}} \,
-
 \frac{ A_4 ({\bm P^\phi_{6}}, {P^\phi_1}, {\bm P^{\eps^M}_{2:4}},5 ) \, A_4 ({\bm P^{\eps^M}_{5:1}}, {\bm 2},3, 4 )   }{s_{2 3 4} }
\right]\,.
 \qquad \nonumber
\end{eqnarray}
%%%%%%%%%%%%%%%%%%%%%%%%%%%%%%
The longitudinal pieces have again cancelled, leaving a simple sum over intermediate polarizations and a very intuitive recursive structure, as shown. Although the explicit evaluation of this expression is straightforward, the resulting expression is lengthy and we do not reproduce it here.
%%%%%%%%%%%%%%%%%%%%%%%%%%%%%%%%%%%
\section{Kleiss-Kuijf decomposition}\label{kkdecomposition}
%%%%%%%%%%%%%%%%%%%%%%%%%%%%%%%%%%%
While the method described in the previous section is straightforward and immediately generalizable to any number of gluons $n$, we wish to point out that an alternative track based on an expansion with analytically computed BCJ-numerators is of comparable simplicity.
The trick is to compute the scattering of two massive scalar fields with massless gluons (eventually gravitons) by 
decomposing the reduced Pffafian in terms of a Kleiss-Kuijf (KK) basis \cite{Kleiss:1988ne} by using 
the Bern-Carrasco-Johansson (BCJ) numerators \cite{Bern:2008qj} for Yang-Mills theory.
This useful technique was developed in\footnote{We thank Y. Geyer for sharing us her Mathematica package allowing us to carry out the master BCJ numerator evaluations in order to rewrite the reduced Pfaffian.} \cite{Cachazo:2013iea,Lam:2016tlk,Fu:2017uzt}.\\[5pt]
Let us recall that our first main goal is to calculate the amplitude 
\be A_n(1_\phi, 2_g, 3_g,\ldots, (n-1)_g, n_\phi)\,, \ee
and in order to avoid dealing with the terms $$\frac{2 P_1\cdot P_n + 2\Delta_{1n}}{\s_{1n}} \ \ {\rm and} \ \ \frac{2 P_n\cdot P_1 + 2\Delta_{n1}}{\s_{n1}}\,,$$ we remove from the reduced Pfaffian the rows/columns $\{ 1,n\}$. Thus, we are looking for the following KK expansion
\begin{eqnarray}
\frac{(-1)^{1+n}}{\s_{1n}}\times {\rm Pf} \left[ (\Psi)^{1n}_{1n}  \right] = \sum_{\rho\in S_{n-2}} N_{(1,\rho(2,\cdots,n-1), n )}\, {\rm PT}(1,\rho(2,\cdots,n-1), n )\,,
\end{eqnarray}
where $N_{(1,\rho(2,\cdots,n-1), n )}$ are the BCJ Yang-Mills numerators and $S_{n-2}$ is the group of the $(n-2)!$ permutations of the set $\{ 2,3,\cdots, n-1\}$. As argued in \cite{Geyer:2017ela}, since the Pfaffian ${\rm Pf} \left[ (\Psi)^{1n}_{1n}  \right]$, is independent of the products $P_1\cdot P_n$ and $\Delta_{1n}=\Delta_{n1}=P_1^2$ the algorithm proposed in ref. \cite{Fu:2017uzt} can be applied. Therefore, the scattering between two massive scalar fields with ($n$-2) gluons can be written as 
%%%%%%%%%%%%%%%%%%%%%%%%%%
\begin{eqnarray}\label{bcjexpansion}
A_n(1_\phi, 2_g, ..., n_\phi) = \!\!\! \sum_{\rho\in S_{n-2}}\!\! m_n[12\cdots n | 1\rho(2\cdots n-1) n] \times N_{(1,\rho(2,\cdots,n-1), n )}\Big|_{\substack{ \eps_1 , \eps_n=(\vec{0},1) }}\,, \qquad
\end{eqnarray}
%%%%%%%%%%%%%%%%%%%%%%%%%%%
where the BCJ numerators $N_{(1,\rho(2,\cdots,n-1), n )}$ can be obtained
by the algorithm developed in \cite{Fu:2017uzt} 
and where $m_n[\a | \b]$ is defined by
%%%%%%%%%%%%%%%%%%%%%%
\begin{equation}\label{massiveBA}
m_n[\a_1\cdots \a_n | \b_1\cdots \b_n] \equiv 
 \int d\mu_n \, {\rm PT}(\a_1, \ldots , \a_n) \times {\rm PT}(\b_1, \ldots , \b_n )\,,
\end{equation}
%%%%%%%%%%%%%%%%%%%%%%
with the massive measure $d\mu_n$ as given in \eqref{measure}.\\[5pt]
To illustrate, let us consider the four-point amplitude $A_4(1_\phi, 2_g, 3_g , 4_\phi)$. From \eqref{bcjexpansion} we arrive at
%%%%%%%%%%%%%%%%%%%%%%%%%
\begin{eqnarray}\label{bcj4pts}
A_4(1_\phi, 2_g, 3_g , 4_\phi) && = N_{(1,2,3, 4 )}\Big|_{\eps_1 , \eps_4=(\vec{0},1) }
 \int d\mu_4 \, {\rm PT}(1,2,3,4) \times  {\rm PT}(1,2,3, 4 ) \nonumber\\
 && +\,
 N_{(1,3,2, 4 )}\Big|_{\eps_1 , \eps_4=(\vec{0},1)  }
 \int d\mu_4 \, {\rm PT}(1,2,3,4) \times  {\rm PT}(1,3,2, 4 )\,. 
\end{eqnarray}
%%%%%%%%%%%%%%%%%%%%%%%%%
Now applying the method of ref. \cite{Fu:2017uzt}, the BCJ numerators are readily found to be given by 
%%%%%%%%%%%%%%%%%%%%%%%%%
\begin{eqnarray}
 N_{(1,2,3, 4 )}\Big|_{\eps_1 , \eps_4=(\vec{0},1) }
\!\!=-2 \, (\eps P)_{ 21 } \, (\eps P)_{ 34 } ,
\quad
 N_{(1,3,2, 4 )}\Big|_{ \eps_1 , \eps_4=(\vec{0},1) }
\!\!= 
2\,(\eps P)_{ 21 } (\eps P)_{ 31 } + 2\,(\eps F P)_{ 231 }\,, \nonumber\\
\end{eqnarray}
%%%%%%%%%%%%%%%%%%%%%%%%%
where we have fixed the reference ordering to be $(1,2,3,4)$. The massive integrals obtained in \eqref{bcj4pts} are straightforward to do using the $\L$-algorithm \cite{Gomez:2016bmv}. We find 
%%%%%%%%%%%%%%%%%%%%%%%%%
\begin{eqnarray}\label{2biadjoints}
&&
m_4[12 34 | 12 34 ] = \int d\mu_4 \, {\rm PT}(1,2,3,4) \times  {\rm PT}(1,2,3, 4 ) 
=\frac{1}{s_{P_12}} + \frac{1}{s_{23}}\,,  
\qquad
 \nonumber\\
 &&
 m_4[12 34 | 13 24 ] =
 \int d\mu_4 \, {\rm PT}(1,2,3,4) \times  {\rm PT}(1,3,2, 4 ) = -\frac{1}{s_{23}}\,.
 \qquad
\end{eqnarray}
%%%%%%%%%%%%%%%%%%%%%%%%%
For the four-point amplitude we therefore get
%%%%%%%%%%%%%%%%%%%%%%%%%
\begin{eqnarray}\label{}
A_4(1_\phi, 2_g, 3_g , 4_\phi) = 
\frac{ -2 \, (\eps P)_{ 21 } \, (\eps P)_{ 34 }  }{ s_{P_12} } + \frac{ -2 \, (\eps P)_{ 21 } \, (\eps P)_{ 34 } -\left( 2\,(\eps P)_{ 21 } (\eps P)_{ 31 } + 2\,(\eps F P)_{ 231 }  \right) }{ s_{23} }\,,
 \nonumber
\end{eqnarray}
%%%%%%%%%%%%%%%%%%%%%%%%%
which agrees with the result we found in equation \eqref{A4pts2}.
%%%%%%%%%%%%%%%%%%%%%%%%%%%%%%%%%%%%
\subsubsection*{Explicit BCJ numerators at five points}
%%%%%%%%%%%%%%%%%%%%%%%%%%%%%%%%%%%%
This method easily generalizes. For two massive scalar legs and three gluons we need to evaluate
%%%%%%%%%%%%%%%%%%%%%%%%%%
\begin{eqnarray}\label{KK5p}
A_5(1_\phi, 2_g, 3_g,4_g, 5_\phi) = \sum_{\rho\in S_{3}}  
m_{5}[12345|1\rho(234)5 ] \times 
N_{(1,\rho(2,3,4), 5 )}\Big|_{ \eps_1 , \eps_5=(\vec{0},1)  } \,. 
\end{eqnarray}
%%%%%%%%%%%%%%%%%%%%%%%%%%%
The six BCJ-numerators $ N_{(1,\rho(2,3,4), 5 )}\Big|_{ \eps_1 , \eps_5=(\vec{0},1)  }$ are given by 
%%%%%%%%%%%%%%%%%%%%%%%%%
\begin{eqnarray}\label{bcj5p}
 &&
\left( \frac{1}{ \sqrt{2} } \right) N_{(1,2,3, 4,5 )}\Big|_{ \eps_1 , \eps_5=(\vec{0},1)  }
\!\!=-2\, 
(\eps P)_{ 21 }\, (\eps P)_{ 45} \, \left[ (\eps P)_{ 31 } + (\eps k)_{ 32 } \right]\,,
\qquad \\ 
&&
\left( \frac{1}{ \sqrt{2}} \right)
N_{(1,4,2, 3,5 )}\Big|_{ \eps_1 , \eps_5=(\vec{0},1)  }
\!\!= 
- 2\,
(\eps P)_{ 25 } \, (\eps P)_{ 41 }
\left[ (\eps P)_{ 21 } + (\eps k)_{ 24 } \right] - 
 (\eps \eps)_{ 34 } (\eps P)_{ 21 } s_{P_14} \nonumber \\
&&
\hspace{4.7cm} 
 +
(\eps \eps)_{ 24 } (\eps P)_{35} s_{P_14}\,, \nonumber \\
&&
\left( \frac{1}{ \sqrt{2}} \right)
N_{(1,3,4,2,5 )}\Big|_{ \eps_1 , \eps_5=(\vec{0},1)  }
\!\!=
 2
 (\eps P)_{ 25 } (\eps P)_{ 31 } \left[ (\eps k)_{ 42 } + (\eps P)_{ 45 } \right] 
 +
( \eps \eps)_{ 23 } \left[ (\eps k)_{ 42 } + (\eps P)_{ 45 } \right] s_{ P_1 3 }  
\nonumber \\
&&
\hspace{4.3cm} 
- (\eps \eps)_{ 43 } (\eps k)_{ 24 } s_{ P_1 3 }  
- (\eps \eps)_{ 24 } (\eps P)_{ 31 } s_{ P_1 34 }  
- 
 (\eps \eps)_{ 24 } \left[ (\eps k)_{ 32 } + (\eps P)_{ 35 } \right] s_{ P_13 } \,
, \nonumber \\
&&
\left( \frac{1}{ \sqrt{2}} \right)
N_{(1,2,4,3,5 )}\Big|_{ \eps_1 , \eps_5=(\vec{0},1)  }
\!\!=
-
2 (\eps P)_{ 21 } (\eps P)_{ 35 }  \left[ (\eps P)_{ 41 } + (\eps k)_{ 42 } \right] - 
  (\eps \eps)_{ 34 } (\eps P)_{ 21 } (s_{ P_1 4 } + s_{ 24 } )\,,
  \nonumber \\
  &&
\left( \frac{1}{ \sqrt{2}} \right)
N_{(1,3,2,4,5 )}\Big|_{ \eps_1 , \eps_5=(\vec{0},1)  }
\!\!=
-2 (\eps P)_{ 31 } (\eps P)_{ 45 } \left[ (\eps P)_{ 21 } + (\eps k)_{ 23 } \right] + 
  (\eps \eps)_{ 23 } (\eps P)_{ 45 } s_{ P_1 3 }\,,
\nonumber \\
  &&
\left( \frac{1}{ \sqrt{2}} \right)
N_{(1,4,3,2,5 )}\Big|_{ \eps_1 , \eps_5=(\vec{0},1)  }
\!\!=
2
(\eps P)_{ 41 } (\eps P)_{ 25 } \left[ (\eps k)_{ 32 } + 
  (\eps P)_{ 35 } \right] - (\eps \eps)_{ 34 } \left[ (\eps P)_{ 21 } + 
   ( \eps k)_{ 23 } \right] s_{ P_1 4 } \nonumber \\
 &&
\hspace{2.2cm}    
   + (\eps \eps)_{ 24 } \left[ (\eps k)_{ 32 } + (\eps P)_{ 35 } \right] s_{ P_1 4 }
  - (\eps \eps)_{ 23 } \left[ (\eps k)_{ 42 } + (\eps P)_{ 45 } \right] s_{ P_14 } - (\eps \eps)_{ 23 } 
  (\eps P)_{ 41 } s_{ P_134 }\,,
\nonumber
\end{eqnarray}
%%%%%%%%%%%%%%%%%%%%%%%%%
where we have fixed the reference ordering to be $(1,2,3,4,5)$.

Using again the $\L$-algorithm \cite{Gomez:2016bmv}, it is straightforward to compute, with two massive legs, 
\begin{eqnarray}\label{bi5p}
&& m_5[12345|12345] = 
\frac{1}{s_{234}\, s_{34}} + \frac{1}{s_{P_12}\, s_{34}} + \frac{1}{s_{23}\, s_{4P_5}} + \frac{1}{s_{4P_5}\, s_{P_12}}
 + \frac{1}{s_{23}\, s_{234}} , \qquad
 \\
&& m_5[12345|14235] = - \frac{1}{s_{23}\, s_{234}}\,, \qquad
\nonumber \\
&& m_5[12345|13425] = - \frac{1}{s_{234}\, s_{34}}\,, \qquad
\qquad \nonumber
 \\
&& m_5[12345|12435] = -
\frac{1}{s_{234}\, s_{34}} - \frac{1}{s_{P_12}\, s_{34}}\,, 
\qquad
\nonumber \\
&& m_5[12345|13245] = 
- \frac{1}{s_{23}\, s_{4P_5}} - \frac{1}{s_{23}\, s_{234}}\,, \qquad
\nonumber \\
&& m_5[12345|14325] = 
\frac{1}{s_{234}\, s_{34}}  
 + \frac{1}{s_{23}\, s_{234}}\,. \qquad
\nonumber
\end{eqnarray}
After substituting eqs. \eqref{bcj5p} and \eqref{bi5p} into \eqref{KK5p} one can check
that the result matches the one given in eq. \eqref{A5expression}.\\[5pt]
This method does have the drawback for $n$ large that
the number of BCJ numerators grow in a factorial way. For instance, to compute the six and seven-point amplitudes one needs to calculate $4!=24$ and $5!=120$ numerators, respectively.  
%%%%%%%%%%%%%%%%%%%%%%
\section{Two massive scalars and gravitons}\label{SGravitons}
%%%%%%%%%%%%%%%%%%%%%%
In the previous sections, we have shown different methods for efficient evaluation of scattering amplitudes of two massive scalar fields and $(n-2)$ gluons. Staying within the CHY-framework as in section \ref{SGluons}, one could similarly express the amplitude of the scattering
among two massive scalars ($\phi$) and gravitons ($h_a$) through \cite{Cachazo:2013gna,Cachazo:2013iea,Cachazo:2015nwa},
%%%%%%%%%%%%%%%%%%%%%%%%
\begin{eqnarray}\label{ChyPrescriptionG}
{\cal M} _n (2\phi,(n-2)h) = \int d\mu_n \,\, {\rm Pf}^\prime \Psi_n\Big|_{ \eps_1 , \eps_n=(\vec{0},1)  } \times {\rm Pf}^\prime \Psi_n\Big|_{ \eps_1 , \eps_n=(\vec{0},1)  } \, ,
\end{eqnarray}
%%%%%%%%%%%%%%%%%%%%%%%%
where the gravitons are identified as, $h_a^{\mu\nu}\equiv \eps^\mu_{a}\eps^\nu_{a}$ and using 
the same massive measure defined in \eqref{measure}. Similarly, one can use a KK-decomposition analogous to what we explained above for the case of gluons in \eqref{bcjexpansion}, and write
%%%%%%%%%%%%%%%%%%%%%%%%
\begin{eqnarray}\label{}
\!\!\!\!\!\!\!\!\!{\cal M} _n (2\phi, (n-2) h) =\!\! \sum_{\rho \in S_{n-2} \atop \delta \in S_{n-2}} \! N_{ (1, \rho, n )}\Big|_{ \eps_1 , \eps_n=(\vec{0},1)  } \times  m_n [1\, \rho\, n |1 \, \delta \, n] \times N_{ (1, \delta,n)}\Big|_{ \eps_1 , \eps_n=(\vec{0},1)  }\,.
\end{eqnarray}
However, by using the Kawai-Lewellen-Tye (KLT) \cite{Kawai:1985xq} relations at the amplitude level, it seems much more straightforward to find the scattering between two massive scalar and $(n-2)$ gravitons by use of the momentum kernel \cite{BjerrumBohr:2010yc,BjerrumBohr:2010hn}, $i.e$,
\begin{eqnarray}\label{Mn2p2ha}
\!\!\!\!\!\!\!\!\!\!{\cal M} _n (2\phi,(n\!-\!2)h)\! =(-1)^{n-3} \sum_{\a \in S_{n\!-\!3} \atop \b \in S_{n\!-\!3}} \!\!\! A_n(1_{\phi}, \a_g, (n\!-\!1)_g,n_\phi) \times \! {\cal S}{[\a|\b]_{k_1}} \! \times \! \\ \vspace{2.5cm} A_n(n_\phi,(n\!-\!1)_g, \b_g,1_{\phi})\,. 
\end{eqnarray}
Here $A_n$ is an amplitude of two massive scalars and $(n-2)$ gluons as defined in \eqref{AnSgluons}, and the momentum kernel ${\cal S}[\alpha|\beta] $ is
\begin{equation}
{\cal S}[i_1,\ldots,i_k | j_1,\ldots, j_k]_{k_1} \equiv \prod_{t=1}^k\left(s_{{i_t} 1}+\sum_{q>t}^k \Theta(i_t,i_q)s_{{i_t},{i_q}}\right),\end{equation}
where $\Theta$ is the step function. 
%%%%%%%%%%%%%%%%%%%%%%%%
For instance, for the four-point amplitude we immediately get 
%%%%%%%%%%%%%%%%%%%%%%%%
\begin{eqnarray}
{\cal M} _4 (2\phi,2 h) = A_4(1_{\phi}, 3_g, 2_g ,4_\phi) \times  {\cal S}{[3|2]} \times A_4(1_{\phi}, 2_g, 3_g, 4_\phi) \,,
\end{eqnarray}
where ${\cal S}{[3|2]} = -s_{23}$,
%%%%%%%%%%%%%%%%%%%%%%%%
thus using the result found in \eqref{A4pts2}, one has
%%%%%%%%%%%%%%%%%%%%%%%%
\begin{eqnarray}
{\cal M} _4 (2\phi,2 h) &=& 
 \left[ \frac{ 2\, (\eps P)_{31} (\eps k)_{23} \, s_{P_13} - 2\, (\eps F P )_{321}\, s_{P_13} - 2 \, (\eps P)_{31} (\eps P)_{24} \, s_{23}}{ s_{P_13}\, s_{23} }
\right]  
\times (-s_{23}) \nonumber \\
&& 
\!\!\!\!\!\!
\times  
 \left[ \frac{ 2\, (\eps P)_{21} (\eps k)_{32} \, s_{P_12} - 2\, (\eps F P )_{231}\, s_{P_12} - 2 \, (\eps P)_{21} (\eps P)_{34} \, s_{23}}{ s_{P_12}\, s_{23} }
\right] 
\nonumber \\
&&
\!\!\!\!\!\!\!\!
=-
\frac{
\left [ 2\, (\eps P)_{ 24 } (\eps P)_{ 31 } s_{ P_12 }  
   + 2\, (\eps P)_{ 21 } (\eps P)_{ 34 } s_{2P_4} + 
   (\eps \eps)_{ 23 } s_{ P_12 } s_{2P_4} \right]^2 }
{  s_{ P_12 } \, s_{ 23 } \, s_{2P_4} } \,,
\nonumber
\end{eqnarray}
%%%%%%%%%%%%%%%%%%%%%%%%
which is the correct 4-point amplitude. Higher order amplitudes follow by KLT-squaring analogously.
%%%%%%%%%%%%%%%%%%%%%%%%%%%%%
\section{Conclusion}
We have presented different methods to compute the tree-level scattering amplitudes of two massive scalars and an in principle
arbitrary number of gravitons in $D$-dimensions. These are the tree-level amplitudes needed to obtain the classical two-body
scattering of two massive objects without spin in general relativity through the use of unitarity. The most economical
method appears to be the one based on a new set of recursive relations that can be derived from the so-called $\Lambda$-algorithm
(or double cover) in the CHY-formalism. In this method one first defines an extension of scattering amplitudes where one external
leg is taken off-shell (defining, effectively, a current in the case of Yang-Mills theory) and then glues off-shell legs 
together by an appropriate polarization sum. We have proven a particular simplification in comparison to the pure Yang-Mills
case when the amplitude contains two massive scalar legs: a sum over longitudinal polarizations cancels exactly. The resulting
amplitude relations for two massive scalars and any number of on-shell gluons thus becomes surprisingly simple.

Although a similar technique can be used to compute amplitudes of two massive scalars with an arbitrary number of gravitons
we have found it economical to simply use KLT-squaring in order to obtain these. Again, they are then provided in
$D$-dimensions and with arbitrary polarization tensors. 

We have checked our general recursive formula up to six points with existing expressions in the literature for the case
$D=4$, always finding complete agreement. An interesting observation is the possibility of establishing a new on-shell
set of recursion relations for these amplitudes based on BCFW-recursion combined with the double-cover analysis of the
$\Lambda$-algorithm. This will be discussed elsewhere.
%%%%%%%%%%%%%%%%%%%%%%%%%
\subsection*{Acknowledgements}
%%%%%%%%%%%%%%%%%%%%%%
We would like to thank Y. Geyer and J. Agerskov for helpful discussions. This project has received funding from the European Union's Horizon 2020 research and innovation programme under the Marie 
Sklodowska-Curie grant agreement No. 764850 ``SAGEX'' and has also been supported in part by the Danish National Research Foundation (DNRF91). H.G. acknowledges partial support from University Santiago de Cali (USC).
\appendix
\section{Higher-point Yang-Mills amplitudes with massive legs}\label{Off-shell-YM}
Here we present details of the main ingredients that go into the computation of the massive five-point gluon amplitudes: $A_5 ({\bm P_1},P_2,{\bm P_3},4,5 ) $, $A_5 ({\bm P_1},2,{\bm P_3}, P_4,5 )$ and $A_5 ({\bm P_1},{\bm P_2}, $ $P_3,4,5 ) $.\\[5pt]
Using the method developed in \cite{Gomez:2018cqg}, the factorization decomposition of $A_5 ({\bm P_1},P_2,{\bm P_3},4,5 )$ becomes 
%%%%%%%%%%%%%%%%%%%%%%%%%%%%%%
\begin{eqnarray}\label{YM5pts}
&&
A_5 ({\bm P_1},P_2,{\bm P_3},4,5 )  
=(-1)\times \sum_M\left\{ \frac{ A_3 ({\bm P^{\eps^M}_{5:2}},{\bm P_3}, {4} ) \times A_4 ({\bm P_1}, P_2, {\bm P^{\eps^M}_{34}}, 5)}{s_{P_34}} \, + \right. 
\nonumber\\
&&
\left. 
 \frac{ A_3 ({\bm P^{\eps^M}_{3:5}},{\bm P_{1}}, {P_2} ) \times A_4 ({\bm P^{\eps^M}_{12}}, {\bm P_3}, 4, 5 )   }{s_{P_345} } +
 \frac{ A_3 ({\bm P_{3}},{\bm P^{\eps^M}_{4:1}}, {P_2} ) \times A_4 ({\bm P_1}, {\bm P^{\eps^M}_{23}} , 4, 5 )   }{s_{45P_1} } \right\}  \nonumber\\
&& 
+ 2 \sum_L \left\{ 
 \frac{ A_3 ({\bm P^{\eps^L}_{513}},{\bm P_2}, {4} ) }{s_{P_24}}  
 \times A_4 ({\bm P_1}, P_3, {\bm P^{\eps^L}_{24}}, 5)
+ \, 
 \frac{ A_4 ({\bm P^{\eps^L}_{13}}, {\bm P_2},4,5 ) }{s_{P_245} } \,\,
 \times
 \right.
 \nonumber
 \\
&&  
\qquad\quad
\left.
 \, A_3 ({\bm P^{\eps^L}_{245}},{\bm P_{1}}, {P_3} ) 
 \right\} ,
 \qquad 
\end{eqnarray}
%%%%%%%%%%%%%%%%%%%%%%%%%%%%%%
where we have written $A_5 ({\bm P_1},P_2,{\bm P_3},4,5 ) $ in terms of the smaller amplitudes, $A_3 ({\bm P_a},{\bm P_b},$ $P_c )$, $A_4 ({\bm P_a},P_b,{\bm P_c},d )$ and $A_4 ({\bm P_a},{\bm P_b}, P_c,d )$. As in the four-point case, we must use the identities in \eqref{Eta} and \eqref{Longitudinal}.\\[5pt]
It is straightforward to find the longitudinal contributions,
%%%%%%%%%%%%%%%%%%%%%%%%%%%%%%
\begin{eqnarray}\label{LA5-one}
&&
- 2 \sum_L \left [
 \frac{ A_3 ({\bm P^{\eps^L}_{513}},{\bm P_2}, {4} ) }{s_{P_24}}  
 \times A_4 ({\bm P_1}, P_3, {\bm P^{\eps^L}_{24}}, 5)
\right] = \nonumber \\
&&
-
 \sqrt{2}\, (\eps_2\cdot \eps_4)\times 
 \frac{  s_{5P_{13}} (\eps_1\cdot F_5\cdot \eps_3)  + 2
 (\eps_1 \cdot \eps_3) 
(P_1\cdot F_5\cdot P_3 )
  }{  s_{5P_{13}} \, s_{P_15} } , 
  \qquad \qquad
\end{eqnarray}
%%%%%%%%%%%%%%%%%%%%%%%%%%%%%%
and 
%%%%%%%%%%%%%%%%%%%%%%%%%%%%%%
\begin{eqnarray}\label{LA5-two}
&&
- 2 \sum_L \left [
 \frac{ A_4 ({\bm P^{\eps^L}_{13}}, {\bm P_2},4,5 ) }{s_{P_245} } 
 \times  
A_3 ({\bm P^{\eps^L}_{245}},{\bm P_{1}}, {P_3} ) 
\right] = 
\nonumber
\\
&&
-
\sqrt{2}\, (\eps_1 \cdot \eps_3) \times
 \frac{ s_{5P_{24}}\, (\eps_2\cdot F_5 \cdot \eps_4)  +2
(\eps_2\cdot \eps_4) (P_2\cdot F_5 \cdot k_4) 
  }{ s_{5 P_{24} } \, s_{45} } .
\end{eqnarray}\\[5pt]
%%%%%%%%%%%%%%%%%%%%%%%%%%%%%%
Similarly, the amplitude $A_5 ({\bm P_1}, 2,{\bm P_3}, P_4,5 )$ is factorized according to
%%%%%%%%%%%%%%%%%%%%%%%%%%%%%%
\begin{eqnarray}\label{YM5pts2}
&&A_5 ({\bm P_1}, 2,{\bm P_3}, P_4,5 )  
= (-1) \times  \sum_M\left\{ \frac{ A_3 ({\bm P^{\eps^M}_{5:2}},{\bm P_3}, {P_4} ) \times A_4 ({\bm P_1}, 2, {\bm P^{\eps^M}_{34}}, 5)}{s_{P_3P_4} + 2\, \Delta_{34} } \, + \right. 
\nonumber\\
&&
\left . 
 \frac{ A_3 ({\bm P^{\eps^M}_{3:5}},{\bm P_{1}}, {2} ) \times A_4 ({\bm P^{\eps^M}_{12}}, {\bm P_3},P_4 ,5 )   }{s_{P_3P_45} + 2\, \Delta_{34}} +
 \frac{ A_3 ({\bm P_{3}},{\bm P^{\eps^M}_{4:1}}, {2} ) \times A_4 ({\bm P_1}, {\bm P^{\eps^M}_{23}} , P_4 ,5 )   }{s_{P_45P_1} + 2\, \Delta_{14} } \right\}  \nonumber\\
&& 
+ 2 \sum_L \left\{ 
 \frac{ A_3 ({\bm P^{\eps^L}_{513}},{\bm 2}, {P_4} ) }{s_{2P_4} + 2\, \Delta_{14} + 2\, \Delta_{34} }  
 \, A_4 ({\bm P_1}, P_3, {\bm P^{\eps^L}_{24}}, 5)
\!\!\!
+ 
 \frac{ A_4 ({\bm P^{\eps^L}_{13}}, {\bm 2},P_4, 5 ) }{s_{2P_45}+ 2\, \Delta_{14} + 2\, \Delta_{34}
 } \,\,
 \times
 \right.
 \nonumber
 \\
&& \left.  
 \qquad \quad A_3 ({\bm P^{\eps^L}_{245}},{\bm P_{1}}, {P_3} )
 \right\} ,
 \qquad 
\end{eqnarray}
%%%%%%%%%%%%%%%%%%%%%%%%%%%%%%
where we have used \eqref{solthree}, namely
\begin{eqnarray}\label{Deltas}
\Delta_{13}= \frac{  P_1^2 +P_3^2 - P_4^2 }{2} , \quad  \Delta_{14}= \frac{P_1^2 -P_3^2 + P_4^2}{2}  , \quad
 \Delta_{34}= \frac{-P_1^2 +P_3^2 + P_4^2}{2} . \qquad\qquad
\end{eqnarray}
%%%%%%%%%%%%%%%%%%%%%%%%
We also recall that $\Delta_{14}+\Delta_{34}=P_4^2$.
It is now straightforward to verify that the longitudinal contributions in \eqref{YM5pts2} are identical to the one evaluated above, \ie, 
\begin{eqnarray}
- 2 \sum_L \left[ 
 \frac{ A_3 ({\bm P^{\eps^L}_{513}},{\bm 2}, {P_4} ) }{s_{P_42} + 2\, \Delta_{14} + 2\, \Delta_{34}} \times
\, A_4 ({\bm P_1}, P_3, {\bm P^{\eps^L}_{24}}, 5)
 \right]=
 \eqref{LA5-one} \nonumber
\end{eqnarray}
and
\begin{eqnarray}
- 2 \sum_L \left[
 \frac{ A_4 ({\bm P^{\eps^L}_{13}}, {\bm 2},P_4, 5 ) }{s_{P_425}+ 2\, \Delta_{14} + 2\, \Delta_{34}} \times
\, \, A_3 ({\bm P^{\eps^L}_{245}},{\bm P_{1}}, {P_3} )
\right]
 =
\eqref{LA5-two} . \nonumber
\end{eqnarray}

Finally, we are able to expand the amplitude, $A_5 ({\bm P_1},{\bm P_2}, P_3, P_4,5 ) $, in terms of the two previous ones, 
$A_5 ({\bm P_a},P_b,{\bm P_{c}}, d,e )$ and $A_5 ({\bm P_a},b, {\bm P_{c}},P_d,,e )$.  
Using the BCJ-like identity \cite{Bjerrum-Bohr:2016juj,Cardona:2016gon},
$$
s_{P_345} \, {\rm PT}(3,4,5,1,2) + (s_{P_345} + s_{P_15} )\, {\rm PT}(3,4,1,5,2)+ (s_{P_345} + s_{P_1P_{45}} )\, {\rm PT}(3,1,4,5,2)=0,
$$
the amplitude, $A_5 ({\bm P_1},{\bm P_2}, P_3, 4,5 )$, turns into 
\begin{eqnarray}
&&
A_5 ({\bm P_1},{\bm P_2}, P_3, 4,5 ) =\nonumber\\
&&
-
 \left(1+ \frac{ s_{P_{1}P_{45}}}{ s_{P_345} } \right) A_5 ({\bm P_2},P_3,{\bm P_{1}}, 4,5 )
 -
 \left( 1+\frac{  s_{P_{12}6}}{ s_{P_345} } \right) A_5 ({\bm P_{1}},5,{\bm P_{2}}, P_3,4 ) . \qquad \qquad \quad
\end{eqnarray}

Finally, let us show how to compute the six-point amplitude, $A_6 ({\bm P_1},P_2,{\bm P_3},4,5,6 ) $.
The factorization decomposition of $A_6 ({\bm P_1},P_2,{\bm P_3},4,5,6 )$ is given by 
%%%%%%%%%%%%%%%%%%%%%%%%%%%%%%
\begin{eqnarray}\label{YM6pts}
&&A_6 ({\bm P_1},P_2,{\bm P_3},4,5,6 )  
=\sum_M\left\{
 \frac{ A_3 ({\bm P_{3}} , {\bm P^{\eps^M}_{4:1}}, {P_2} ) \times  A_5 ({\bm P_1},{\bm P^{\eps^M}_{23}},4, 5,6)  }{s_{456P_1} }
\, + \right.
\nonumber\\
&&
\quad\quad
 \frac{ A_3 ({\bm P^{\eps^M}_{3:6}},{\bm P_{1}}, {P_2} ) \times A_5 ({\bm P^{\eps^M}_{12}},{\bm P_3},4, 5,6 ) }{s_{P_3456} } 
 +
 \frac{ A_3 ({\bm P^{\eps^M}_{5:2}},{\bm P_3}, {4} ) \times A_5 ({\bm P_1}, P_2, {\bm P^{\eps^M}_{34}}, 5,6)}{s_{P_34}} \,
 \nonumber\\
&& 
\quad\quad
\left.
-
 \frac{ A_4 ({\bm P_{1}}, {P_2}, {\bm P^{\eps^M}_{3:5}},6 ) \times A_4 ({\bm P^{\eps^M}_{6:2}}, {\bm P_3},4, 5 )   }{s_{P_345} }
\right\}
 \nonumber\\
&& 
- 2
 \sum_L
 \left\{ 
 \frac{ A_3 ({\bm P^{\eps^L}_{3:6}},{\bm P_{1}}, {P_2} ) \times A_5 ({\bm P^{\eps^L}_{12}},{\bm P_3},4, 5,6 ) }{s_{P_3456} } 
 +
 \frac{ A_3 ({\bm P^{\eps^L}_{5:2}},{\bm P_3}, {4} ) \times A_5 ({\bm P_1}, P_2, {\bm P^{\eps^L}_{34}}, 5,6)}{s_{P_34}} \,
 \right.
\nonumber\\
&& 
\quad\quad
\left.\left.
-
 \frac{ A_4 ({\bm P_{1}}, {P_2}, {\bm P^{\eps^L}_{3:5}},6 ) \times A_4 ({\bm P^{\eps^L}_{6:2}}, {\bm P_3},4, 5 )   }{s_{P_345} }
\right\} \right|_{
\substack{ \hspace{-1.7cm} 2\leftrightarrow 3 \\ (\eps_\a \cdot P_{A}) = - (\eps_\a \cdot P_{\bar{A}}) \\ \hspace{-1.1cm} (\eps_\a\cdot P_\a)=0 }
  },
 \qquad 
\end{eqnarray}
%%%%%%%%%%%%%%%%%%%%%%%%%%%%%%
where $2 \leftrightarrow 3$ means the changing of the two labels, $\a=1,3$ and $P_{\bar{A}}$ is the complement of $P_{A}$ (by the momentum conservation condition, $P_{{A}}+P_{\bar{A}}=0$). For example, $P_{A}$ is given by $P_{2456}, P_{24}$ and $P_{245} $  
in the last three term in \eqref{YM6pts}, respectively, therefore, $P_{\bar{A}}$ is $P_{13}, P_{1356}$ and $P_{136}$.
Additionally, the identities in \eqref{Eta} and \eqref{Longitudinal} must be used in the above factorization expansion.
%%%%%%%%%%%%%%%%%%%%%%%%%%%%%%%%%%%%%%%%%%%%
%%%%%%%%%%%%%%%%%%%%%%%%%%%%%%%%%%%%%%%%%%
\section{Longitudinal contributions}\label{preposition}
%%%%%%%%%%%%%%%%%%%%%%%%%%%%%%%%%%%%%%%%%%
As we have observed in all special cases worked out in this paper, the longitudinal contributions to the factorized amplitudes with massive scalars always vanish identically. In this section we prove this important fact in all generality.\\[5pt]
Let us consider a Yang-Mills $n$-point amplitude with up to three massive legs $A_n ({\bm P_n},P_1,$ ${\bm P_2},3,..., n-1 ) $.
Applying the factorization method, a generic longitudinal contribution is given by 
%%%%%%%%%%%%%%%%%%%%%
\begin{eqnarray}
\sum_{L}
\frac{ A_{(n-i)+2}( {\bm P_n}, { P_2}, {\bm P^{\eps^L}_{134...i} },i+1,...,n-1) \times A_i( {\bm P^{\eps^L}_{i+1... n 2} } , {\bm P_1}, 3,4, ..., i) }{ s_{P_134...i} } , \qquad
\end{eqnarray}
%%%%%%%%%%%%%%%%%%%%%
where the the two amplitudes are sewn together by the rule 
\begin{equation}
\sum_L \eps^{L\, \mu}_{i} \eps^{L\, \nu}_{j} = \frac{P^\mu_{i}\, P^\nu_{j} }{P_{i}\cdot P_{j} + P_n^2 - P_2^2} ~. 
\end{equation}
We can now show the following: \\[5pt]
Under the condition $\eps_1=\eps_n=(\vec{0},1)$, the amplitudes, $A_{(n-i)+2}( {\bm P_n}, { P_2}, {\bm P^{\eps^L}_{134...i} }, i+1,...,n-1)$ and $ A_i( {\bm P^{\eps^L}_{i+1... n 2} } , {\bm P_1}, 3,4, ..., i)$ vanish identically.
\\[5pt]
The proof of this proposition is straightforward. \\[5pt] Let us consider the amplitude, $ A_i( {\bm P^{\eps^L}_{i+1... n 2} } , {\bm P_1}, 3,4, ..., i)$. From the notation introduced in the main text, it is clear that the reduced matrix $\left[ (\Psi_i)^{P_{i+1... n 2} P_1}_{P_{i+1... n 2} P_1} \right]$, 
has a row (column) given by the vector
%%%%%%%%%%%%%%%%%%%%%%%%%%%%%%%%%%%%
\begin{eqnarray}
\left(
\begin{matrix}
\frac{\eps_1 \cdot k_2 } {\s_{12}} , &
\cdots , &
\frac{ \eps_1 \cdot k_i} {\s_{1i} } , &
\frac{ \eps_1 \cdot \eps^L_{i+1...n2} } {\s_{1 P_{i+1...n2}} } ,&
0 ,&
\frac{ \eps_1 \cdot \eps_2} {\s_{12} } , 
\cdots , &
\frac{ \eps_1 \cdot \eps_i} {\s_{1i} }
\end{matrix}
\right)\Big|_{\substack{ \eps_1,\eps_n=(\vec{0},1) }} \!\!=
\left(
\begin{matrix}
0, &
\cdots , &
0, &
\frac{ \eps_1 \cdot \eps^L_{i+1...n2} } {\s_{1 P_{i+1...n2}} } ,&
0 ,
\cdots , &
0
\end{matrix}
\right). \qquad
\nonumber
\end{eqnarray}
%%%%%%%%%%%%%%%%%%%%%%%%%%%%%%%%%%%
Since $\eps^L_{i+1...n2} $ is proportional to $P_{i+1...n2} = k_{i+1}+\cdots + k_n+P_2$ it follows that 
$$
\frac{ \eps_1 \cdot \eps^L_{i+1...n2} } {\s_{1 P_{i+1...n2}} }  \propto  \frac{ \eps_1 \cdot P_{i+1...n2} } {\s_{1 P_{i+1...n2}} } =0, 
$$
using that $(\eps_1\cdot k_i)=0$. Therefore, $ A_i( {\bm P^{\eps^L}_{i+1... n 2} } , {\bm P_1}, 3,4, ..., i)$ vanishes trivially for, $\eps_1=\eps_n=(\vec{0},1)$. The essential property that makes these contributions vanish is the fact that
the polarization vectors associated with what become massive scalars live in a higher dimensional space with no overlap with the momenta of the $D$-dimensional space.\\[5pt]
The same argument works for $A_{(n-i)+2}( {\bm P_n}, { P_2}, {\bm P^{\eps^L}_{134...i} }, i+1,...,n-1).$

%\bibliographystyle{physics}
%\bibliography{amplitude_refs}
%\end{document}
%\providecommand{\href}[2]{#2}\begingroup\raggedright\begin{thebibliography}{10}
%%%%%%%%%%%%%%
%\bibliographystyle{JHEP}
%\bibliography{mybib}
%
%\bibliographystyle{JHEP}
%\bibliography{mybib}
%\end{thebibliography}\endgroup

\end{document}